\documentclass[twocolumn,showpacs,preprintnumbers,amsmath,amssymb]{revtex4}

\usepackage{epsf}
\usepackage{rotating}

\usepackage{graphicx,epsfig}
\usepackage{dcolumn}
\usepackage{bm}

\def\cN{{\cal N}}

\def\cH{{\cal H}}
\def\cM{{\cal M}}

\def\flow{{I}}
\newcommand{\req}[1]{Eq.~(\ref{#1})}
\newcommand{\avg}[1]{\langle #1\rangle}

\newcommand{\fig}[1]{Fig.~\ref{#1}}

\DeclareMathOperator*{\argmin}{{\rm argmin}}	

\newcommand{\cut}[1]{{}}
\newcommand{\etal}[1]{\emph{~et al.}}
\begin{document}

\preprint{}

\title[Title]
{The Competition for Shortest Paths on Sparse Graphs}
\author{Chi Ho Yeung and David Saad}
\affiliation{The Nonlinearity and Complexity Research Group, Aston University, Birmingham B4 7ET, United Kingdom}

\date{\today}

\begin{abstract}
Optimal paths connecting randomly selected network nodes and fixed routers are studied analytically in the presence of non-linear overlap cost that penalizes congestion. Routing becomes increasingly more difficult as the number of selected nodes increases and exhibits ergodicity breaking in the case of multiple routers. A distributed linearly-scalable routing algorithm is devised. The ground state of such systems reveals non-monotonic complex behaviors in both average path-length and algorithmic convergence, depending on the network topology, and densities of communicating nodes and routers.
\end{abstract}

\maketitle


Routing and path selection are at the heart of many communication and logistics applications. For instance, instant messengers, Internet telephony and payment security verification require packets to be delivered instantly or otherwise lose their functionality~\cite{huitema95, moy98}; while the efficiency of transportation networks depends crucially on effective path selection~\cite{wu08, kim02}. Existing routing algorithms are mostly based on minimizing path lengths. Some use routing tables that register the shortest distance to various destinations but are insensitive to traffic congestion~\cite{bellman58, dijkstra59}; others control congestion by monitoring queue length or latency heuristically~\cite{rangwala06}, or merely optimize routing selfishly~\cite{roughgarden02}. Devising efficient distributive principled routing algorithms which minimze route length while restricting congestion remains a challenge.

Path optimality and congestion control have been extensively studied within the physics community in other contexts, such as the research of spanning~\cite{noh02, dobrin01} and Stenier trees~\cite{bayati08} with \emph{quenched} link weights, to mimic broadcast or multi-cast systems. However, these studies ignore interaction terms (overlap costs) that \emph{depend on the specific choice of paths}. Other approaches such as preferential random walk and diffusion methods are used to reduce traffic congestion, but result in heuristic protocols which route packets through sub-optimal paths in a probabilistic manner~\cite{adamic01, wang06, danila06}.

Mapped onto a statistical physics framework, routing poses both theoretical and numerical challenges due to the multiplicity of possible routes between communicating nodes and the nonlinear costs induced by the interaction between overlapping routes, akin to a non-local repulsion force. We remark that although overlap costs have been partially addressed by assigning quenched link weights~\cite{noh02}, they do not fully reflect the complex interaction between dynamical variables. Techniques used to analyze polymers~\cite{daoud75}, for instance of self-avoiding walks~\cite{stilck96} and the traveling salesman problem~\cite{mezard86}, are prime candidates for the analysis of routing problems but do not consider the cost of interaction between paths.

In this Letter we study a scenario whereby numerous senders seek the shortest possible route to a few receivers while minimizing traffic congestion. The problem is relevant to node pairs on a network that communicate via designated routers; or nodes, possibly sensors, that communicate via an outlet router or base stations. It is also relevant to transportation networks where traffic gravitates towards one or several centers, such as city center or hub airports~\cite{wu08, wuellner10}. Using the cavity approach~\cite{mezard87}, we examine analytically the dependence of the typical path length on the system topology, the density of communicating node-pairs and routers and their location. We identify the conditions for ergodicity breaking in solution space and observe oscillations in typical path lengths and algorithmic convergence in regular graphs. We show that allocating routers on hubs, which seems to be the natural choice, is indeed optimal in many respects. An applicable linearly-scalable local message passing algorithm is derived which optimizes, in a principled manner, individual routes subject to the mitigation of global congestion.

\begin{figure}
\leftline{\epsfig{figure=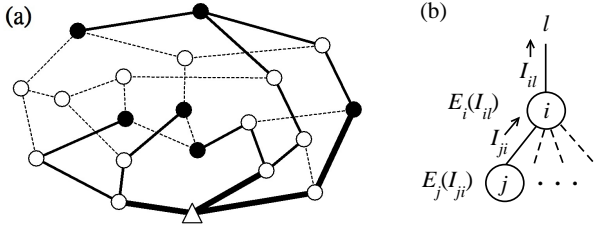, width=0.66\linewidth}
\epsfig{figure=cavity.eps, bb=48 -30 558 644, clip=, width=0.32\linewidth}}
\vspace{-0.5cm}
\caption{
(a) An example of the ground state of $\cH$ with $\alpha\!=\!2$ in a simple network with 6 senders ({\LARGE\textbullet}) and a single receiver ($\triangle$) among 20 nodes.  Dashed lines correspond to idle links, while thin and thick solid lines correspond to communication loads of 1 and 2, respectively. To minimize $\cH$, communication from the top left sender is routed through a long path. (b) a schematic diagram representing the derivation of \req{eq_recur}.
}
\label{fig_ex}
\end{figure}

\emph{Model:} We consider a sparse network of $N$ nodes $i\!=\!1,\ldots, N$ where each node $i$ is randomly connected to $k_i$ neighbors denoted by $\cN_i$; where $k_i\!\ll\!N$ is randomly drawn from a distribution $\rho(k)$. We randomly select one node as a \emph{receiver} (router) denoted by $r$~\cite{footnote}, and a fraction $f_s$ of the other nodes, with $0\le\! f_s\!\le\! 1$, as \emph{senders} that communicate with $r$ through a single path each.
We assign $\sigma_{ij}^{s\to r}\!=\!1$ if the communication originated from $s$ to $r$ passes through the edge from $i$ to $j$, and $\sigma_{ij}^{s\to r}\!=\!0$ otherwise. We also denote the \emph{integer} communication flow from $i$ to $j$ as $\flow_{ij}\!\equiv\!\sum_s\left[\sigma_{ij}^{s\to r}\!-\!\sigma_{ji}^{s\to r}\right]$. To facilitate the analysis, each node $i$ is assigned with a communication load $\Lambda_i$, reflecting its role: the receiver node $r$ is assigned with $\Lambda_r\!=\!-\!\infty$, senders are assigned with $\Lambda\!=\!+\!1$, all other nodes are assigned with $\Lambda\!=\!0$.  Flow on the network is then re-directed until excess communication load $R_i=\Lambda_i\!+\!\sum_{j}A_{ji}\flow_{ji}$ vanishes, i.e. $R_i\!=\!0,~ \forall i\!\neq\!r$. To minimize path length subject to a nonlinear cost for overlapping paths, we introduce the Hamiltonian  $\cH\!=\!\sum_{(ij)}|\flow_{ij}|^\alpha$ with $\alpha\!>\!1$; analysis and algorithm are generic for any $\alpha\!\ge\! 1$. An example of a small system attaining the ground state of $\cH$ with $\alpha\!=\!2$ is shown in \fig{fig_ex}(a).

This model can be adapted to accommodate various routing scenarios. For instance, senders with a positive integer $\Lambda$ may be introduced to model data segmentation and routing through multiple paths; multiple receivers may be considered as is the case in wireless sensor networks where \emph{any} reachable base station will do~\cite{rangwala06}.
The Hamiltonian can also be modified to accommodate different forms of cost and objectives on nodes and links.

To analyze the system's ground state behavior, we employ the cavity approach~\cite{mezard87} and assume that only large loops exist, so that neighbors of node $i$ become statistically independent if it is being removed. This allows one to define $E_i(\flow_{il})$, the energy of a tree terminated at the link from vertex $i$ to parent $l$, given the flow $\flow_{il}$~\cite{wong06}. For any $i\!\neq\! r$, one obtains a recursion equation relating the energy $E_i(\flow_{il})$ to $E_j(\flow_{ji})$ of its neighbors (decedents) $j$ other than (parent) $l$, as sketched in \fig{fig_ex}(b), given by
\begin{eqnarray}
\label{eq_recur}
E_i(\flow_{il}) = \hspace{-0.3cm}
\min_{\left\{ \left\{\flow_{ji}\right\}\left| R_i=0\right\}\right.}
\hspace{-0.2cm}\left[ |I_{il}|^\alpha\!+\!\sum_{j\in\cN_i\backslash l}E_j(\flow_{ji}) \right]\hspace{-0.15cm}.
\end{eqnarray}
For receiver nodes, being universal sinks, $E_r \!=\! |I_{jr}|^\alpha,\forall j\!\in\!\cN_r$. For brevity we denote the right hand side of \req{eq_recur} by a functional $\cM[\underline{E}_{~il}; \Lambda_i, \flow_{il}]$ where $\underline{E}_{~il}\!\equiv\!\left\{E_{j}\left| j\in\cN_i\backslash l\right.\right\}$.
The function $E(I)$ is extensive and difficult to iterate, and is replaced by the intensive quantity $E^V(I) \!\equiv\! E(I)\!-\!E(0)$~\cite{yeung10} to derive the recursive relation
\begin{equation}
\label{eq_recur2}
E^V_i(\flow_{il}) = \cM[\underline{E}^V_{~il}; \Lambda_i, \flow_{il}]
-\cM[\underline{E}^V_{~il}; \Lambda_i, 0].
\end{equation}

To evaluate various physical quantities one employs population dynamics to iterate \req{eq_recur} and obtain a stable distribution $P[ E^V(\flow) ]$. The \emph{ground state energy} is evaluated~\cite{mezard02} by the average energy of an additional node and link; denoting  $\underline{E}^V_{~i}\!\equiv\!\left\{E^V_{j}\left| j\in\cN_i\right.\right\}$, this is given by
\begin{align}
\label{eq_enode}
\avg{E_{\rm node}} &= \avg{ \cM[\underline{E}^V_{~i}; \Lambda, 0]}_{\Lambda, k, \underline{E}^V_{~i}},
\\
\label{eq_elink}
\avg{E_{\rm link}} &= \avg{\min_\flow[E^V_{j_1}(\flow)+E^V_{j_2}(-\flow)-|\flow|^\alpha]}_{E^V_{j_1}, E^V_{j_2}}
\end{align}
averaged over the sender and degree distributions $p(\Lambda)$, $\rho(k)$ and $P[E^V(\flow)]$.
The energy per transmission is thus given by $\avg{E} = (\avg{E_{\rm node}}-\frac{\avg{k}}{2}\avg{E_{\rm link}})/f_s$~\cite{mezard02}.

Obtaining typical path length for non-integer $\alpha$ follows directly from \req{eq_elink} in the form of $\avg{L}=\avg{|\flow^*|}/f_s$, where $\flow^* = \argmin_{\flow}[E^V_{j_1}(\flow)+E^V_{j_2}(-\flow)-|\flow|^\alpha]$, since $\avg{|\flow^*|}/f_s$ is effectively the link occupancy per communication. The case of integer $\alpha$ suffers from degeneracy of $\flow^*$ and is best calculated by propagating, parallel to $E^V$, the function
\begin{eqnarray}
\label{eq_L}
L^V_i(\flow_{il}) = |\flow_{ji}|+\sum_{j\in\cN_i\backslash l} L^V_j[\flow_{ji}^*(\flow_{il})],
\end{eqnarray}
where the optimal flow $\flow_{ji}^*(\flow_{il})$, given $\flow_{il}$, is obtained from \req{eq_recur2}.
While $E^V$ provides the optimal energy, $L^V$ evaluates typical traffic in the optimal state. Average path length per communication is obtained using the distribution $P(E^V, L^V)$ by $\avg{L}\!=\!(\avg{L_{\rm node}}\!-\!\frac{\avg{k}}{2}\avg{L_{\rm link}})/f_s$, where $\avg{L_{\rm node}}$ and $\avg{L_{\rm link}}$ are evaluated similarly to \req{eq_L} via the corresponding $I^*$ flow obtained in Eqs. (\ref{eq_enode}-\ref{eq_elink}).

\begin{figure}
\centerline{\epsfig{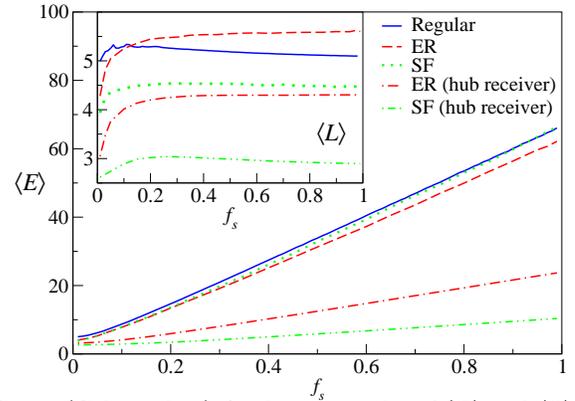}}
\vspace{-0.5cm}
\caption{
(Color online) Analytical results of $\avg{E}$ and $\avg{L}$ per communication as a function of $f_s$ on regular, Erd\"os-R\'enyi (ER) and scale-free (SF) networks, and ER and SF networks with hub receivers; all with $\avg{k}=3$ and $f_r=0.01$.
}
\label{fig_el}
\end{figure}

\emph{Solutions:} Results shown in \fig{fig_el}, for $\alpha\!=\!2$, $f_r\!=\!0.01$~\cite{footnote} and different connectivity distributions representing regular, Erd\"os-R\'enyi (ER) and scale-free (SF) networks, show a linear increase in $\avg{E}$ for most of the range of $f_s$ regardless of network topology. This indicates that the potential quadratic increase in cost due to overlaps is mitigated through rerouting, balancing path-length and overlap costs. Typical path length $\avg{L}$ (inset) increases initially with $f_s$ for all networks (SF being shorter) due to overlap costs, but saturates for ER networks while showing a \emph{decrease} for regular and SF graphs at high $f_s$. This indicates longer alternative routes are also congested, making shorter congested ones more cost-effective. This suggests that simple shortest path routing schemes may achieve close-to-optimal performance for either low or high traffic densities in regular and SF networks, but are less effective for intermediate density and ER networks.

To examine the effect of receiver connectivity, we study the case where receivers are deployed to the largest hubs in ER and SF networks. Figure~\ref{fig_el} shows a more moderate increase in energy and path length compared to random deployment. In addition, communication costs in SF networks become lower with respect to ER networks, reversing their random case positions. This suggests that receiver local connectivity dominates network performance and potentially explains why most transportation networks are SF with hub receivers.

\begin{figure}
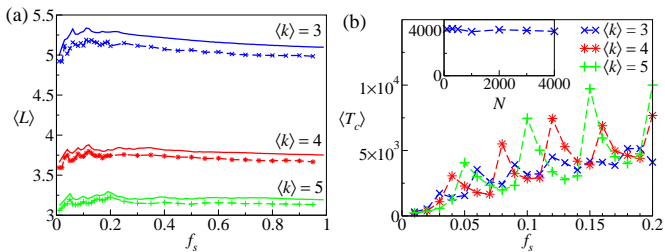

\leftline{\epsfig{figure=path_sim.eps, width=0.5\linewidth} \epsfig{figure=convergenceTime2.eps, width=0.5\linewidth}}
\vspace{-0.2cm}
\caption{
(Color online) (a) Analytic $\avg{L}$ values (solid lines) compared to simulation results (dashed lines, symbols) for regular networks with $\avg{k}\!=\!3,4,5$, a single receiver and $N\!=\!100$ nodes, averaged over 1000 realizations. (b) Algorithmic convergence time $\avg{T_c}$ averaged over instances ($\approx 98\%$) which converge within $5\times 10^4$ updates per node. Inset: $\avg{T_c}$ as a function of $N$ for $\avg{k}\!=\!3$, $f_s\!=\!0.2$ and a single receiver.
}
\label{fig_regular}
\end{figure}

Careful examination of average path length $\avg{L}$ in regular networks as a function of $f_s$ reveals a fine structure of small peaks, in addition to the main peak, at multiples of $\avg{k}/N$ as shown in \fig{fig_regular}(a). These occur when traffic is balanced predominantly around the receiver, which we term the \emph{balanced receiver phenomenon} (BRP). For example, the long route from the top left sender in \fig{fig_ex}(a) is chosen a to ensure that links connecting the receiver are occupied exactly by 2 communications as a high cost is incurred by imbalanced traffic.
This behavior is also reflected in the expected algorithmic convergence time (number of update steps per node) $\avg{T_c}$ shown in \fig{fig_regular}(b). BRP may appear in real-life networks; for instance, one may attempt to balance car traffic heading towards a city center by varying toll or speed limits. This phenomenon is masked in the results for ER and SF networks due to the degree variability of receivers and since BRP is more pronounced at low degree receivers.

\emph{Computation:} One computational challenge in \req{eq_recur2} is the extremization on the integer domain subject to an equality constraint. We use the convexity of $|\flow|^\alpha$ ($\alpha\!\ge\! 1$) to show that $E^V(\flow)$ is convex and denote the energy change when $\flow_{il}$ increases/decreases by 1 as $\Delta_i^{\pm}(\flow_{il}) \!=\! E^V_i(\flow_{il}\!\pm\!1)\!-\!E^V_i(\flow_{il})$. The optimality condition
$\Delta_{j_1}^+[\flow^*_{j_1i}(\flow_{il})]\!+\!\Delta_{j_2}^-[\flow^*_{j_2i}(\flow_{il})]\!\ge \! 0,~\forall j_1, j_2\!\in\!\cN_i\backslash l$,
and the convexity of $E^V_i(\flow_{il})$ yield
\begin{align}
\label{eq_simDelta}
\hspace{-0.2cm}
&\Delta_i^\pm(\flow_{il}) = |\flow_{il}\pm 1|^\alpha- |\flow_{il}|^\alpha +\min_{j\in\cN_i\backslash l}\{\Delta_j^\pm[\flow^*_{ji}(\flow_{il})]\}
\\
\hspace{-0.2cm}
&\flow^*_{ji}(\flow_{il}\pm 1) =
\begin{cases}
\flow^*_{ji}(\flow_{il})\pm 1, & \hspace{-0.15cm}j=\argmin\limits_{j\in\cN_i\backslash l}\{\Delta_j^\pm[\flow^*_{ji}(\flow_{il})]\}
\\
\flow^*_{ji}(\flow_{il}), & \hspace{-0.15cm}\mbox{otherwise}
\end{cases} \nonumber
\end{align}
This simplifies the computation (see Supplemental Material) and facilitates a study of the model's ergodicity breaking properties. Equation~(\ref{eq_simDelta}) implicitly assumes localized interdependence of flows that corresponds to a \emph{replica symmetry} (RS)~\cite{mezard87} property. Small variations in \emph{cavity field} $\Delta_i^\pm$ in \req{eq_simDelta} take the form $\delta\Delta_i^\pm(\flow_{il})\!=\!\delta\Delta_{j^*}^\pm(\flow_{j^*i})$, where $j^*\!=\!\argmin_{j\in\cN_i/l}[\Delta_{j}^\pm(\flow_{ji})]$, leading to expected perturbations of the form $\avg{(\delta\Delta_i^\pm)^2}\!=\!f_s\avg{(\delta\Delta_j^\pm)^2}$ for all system nodes. Perturbations decay for $f_s\!<\!1$ indicating an RS~\cite{rivoire04} behavior, which leads to a unique state. This is supported by simulations as almost all cases converge in a small number of updates.

\emph{Algorithm:} For optimizing individual instances one iterates \req{eq_recur2} until convergence of the $E^V$'s, and computes the flows using either Eqs.~(\ref{eq_enode}) or (\ref{eq_elink}) to facilitate the identification of individual paths. The situation becomes more complicated when solutions of Eqs.~(\ref{eq_enode}) or (\ref{eq_elink}) have degeneracy, which is typical for integer $\alpha$. One can then break the degeneracy by assigning randomly to each link $(ij)$ a small quenched bias $\epsilon_{(ij)}$, and adopts in the iteration of \req{eq_recur2} the modified cost of $|\flow_{ij}|^\alpha\!+\!|\flow_{ij}|\epsilon_{(ij)}$. The iteration then converges to a particular solution with no degeneracy. The algorithm is linearly scalable as shown in the inset of \fig{fig_regular}(b), but the bias increases convergence time with respect to having a non-zero bias.

\emph{Breaking of ergodicity:} While \emph{replica symmetry breaking} (RSB)~\cite{mezard87} which is associated with algorithmic hardness does not emerge in the original model, it does emerge in very similar scenarios. One such case is that of multiple receiver classes, where each sender is assigned a specific receiver type. This scenario corresponds to users/sensors with specific information for specific routers/receivers. In the case of two different receiver types, say A and B with $(\Lambda^A, \Lambda^B)\!=\!(\!-\!\infty, 0)$ and $(0, \!-\!\infty)$, respectively, senders are initialized with either $(\Lambda^A, \Lambda^B)\!=\!(\!+\!1, 0)$ or $(0, \!+\!1)$. The corresponding Hamiltonian $\cH\!=\!\sum_{(ij)}(|\flow_{ij}^A|\!+\!|\flow_{ij}^B|)^\alpha$ and $E^V(\flow_{ij}^A, \flow_{ij}^B)$'s are two-dimensional; $E^V$'s are no longer convex for the entire range and \req{eq_simDelta} does not hold. Nevertheless, one can derive an approximate recursion relation based on the insight gained from \req{eq_simDelta}, for instance
\begin{align}
\label{eq_2classDelta}
\Delta_i^{A^+}& (\flow_{il}^A, \flow_{il}^B) \approx (|\flow_{il}^A+1|+|\flow_{il}^B|)^\alpha - (|\flow_{il}^A|+|\flow_{il}^B|)^\alpha
\nonumber\\
&+ \min\left\{\min_{j\in\cN\backslash l}[\Delta_j^{A^+}], \min_{j_1, j_2\in\cN\backslash l}[\Delta_{j_1}^{A^{2+}}+\Delta_{j_2}^{A^-}], \right.
\nonumber\\
&\left.\min_{j_1, j_2\in\cN\backslash l}[\Delta_{j_1}^{A^+, B^\pm}+\Delta_{j_2}^{B^\mp}]    \right\}~.
\end{align}
Superscripts $A^+$ and $A^{2+}$ correspond to the increase of type A flow by 1 and 2 units, respectively (similarly for B). Various combinations of $\Delta_j$'s on the right hand side add up to give $\Delta_i^{A^+}$; the arguments of $\Delta_j$ were omitted for clarity. This approximation provides good results compared to the exact computation and can be improved by adding combinations on the right hand side.

We employ \req{eq_2classDelta} to identify the RS-RSB transition by introducing perturbation to the cavity fields. The expression becomes more involved and perturbations depend on perturbations induced by a number of descendent (see Supplemental Material). Depending on system parameters, we identify decaying perturbations with RS behavior and diverging perturbations with RSB.

\begin{figure}[t]
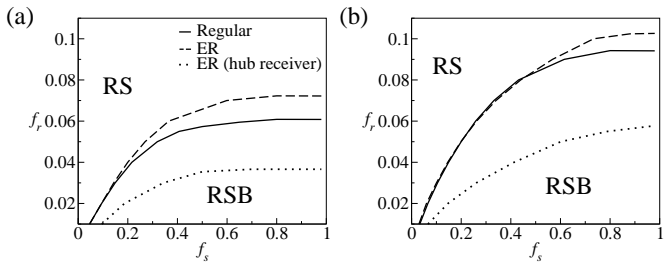

\leftline{\epsfig{figure=phase_fixed.eps, width=0.5\linewidth} \epsfig{figure=phase_flexi.eps, width=0.5\linewidth}}
\vspace{-0.2cm}
\caption{
Phase diagram for fractions $f_s$ and $f_r$ of sender and receiver (router) nodes, respectively, and two types of receivers on regular networks, and ER networks with random and hub receivers, for (a) senders with a fixed router allocation and (b) sender pairs with flexible router choice.
}
\label{fig_phase}
\end{figure}
Figure \ref{fig_phase}(a) shows the RS and RSB phase transitions as a function of the fractions $f_s$ and $f_r$ of sender and router nodes with respect to the system size, respectively, for regular networks, and ER networks with random and hub routers ($f_r$ refers equally to each of the receiver types). Generally, RSB emerges for large $f_s$ and small $f_r$ as a large number of communications from distant senders are optimized simultaneously, leading to an extensive frustration of a finite fraction of all communications. While ER networks with random receiver have a larger RSB phase than the regular network, it shrinks significantly when receivers are allocated on the hubs. This shows again the advantage of hub receivers since the RS phase, where simple algorithms suffice, extends to larger systems with denser communications. Since the allocation of specified receivers to senders induces RSB we expect a more extensive RSB phase with the increase in receiver types. This is relevant to \emph{peer-to-peer} (P2P) networks where each node represents a different receiver type.

In another senario, senders pair up to establish a communication line via a receiver, which can be of either type A or B, by setting their respective communication loads $(\Lambda^A, \Lambda^B)$ to either $(1,0)$ or $(0,1)$. One thus introduces a new variable representing the energy change when sender $i$ switches from receiver A to B
\begin{align}
\phi_i^{A^+, B^-}= &\cM[\underline{E}^V_{i};\Lambda^A\!=\!1,\Lambda^B\!=\!0;0]
\nonumber\\
&-\cM[\underline{E}^V_{i};\Lambda^A\!=\!0,\Lambda^B\!=\!1;0]~,
\end{align}
and its distribution $P(\phi^{A^+, B^-})$. This impacts on recursion equation for $E^V$ which translates to adding a term $\delta_{\Lambda^A(\flow_{il}^A, \flow_{il}^B), 1}\min_{j\in\cN\backslash l}[-\phi^{A^+, B^-}+\Delta_j^{A^{2+}, B^-}]+\delta_{\Lambda^B(\flow_{il}^A, \flow_{il}^B), 1}\min_{j\in\cN\backslash l}[\phi^{A^+, B^-}+\Delta_j^{B^+}]$ to the argument of the outermost minimum function in \req{eq_2classDelta}. The first and second term here correspond to the switch from receiver A to B and from B to A, respectively.

Identifying the RS-RSB transition is carried out in a similar manner to the previous case but one has to account also for the perturbation $\delta\phi^{A^+, B^-}$, resulting in the qualitatively similar phase diagram shown in \fig{fig_phase}(b).

Optimal routing in the presence of overlap costs is highly relevant to communication networks and exhibits a complex and interesting behavior as a function of system topology, sender and receiver densities and types. Results, obtained analytically and verified numerically, provide insight into the suitability of various network topologies as communication networks and the preferred location of network servers. While the original framework is replica symmetric, related scenarios whereby senders communicate individually, or jointly as a pair, through \emph{specific} receivers exhibit RSB behavior, with direct impact on the algorithmic hardness of the routing problem. A generic linearly scalable algorithm which optimizes paths to central routers was also devised. The application of statistical mechanics to communication networks is highly promising both due to the insight gained and the promise of better principled routing algorithms.

We thank K.~Y.~M.~Wong, R.~Urbanke, C.~Fraguoli, N. Ruozzi and A. Mozeika for fruitful discussions. This work is supported by EU FET project STAMINA (FP7-265496) and Royal Society Exchange Grant IE110151.


\newpage
\clearpage
\setcounter{figure}{0}
\setcounter{table}{0}
\setcounter{equation}{0}
\setcounter{page}{1}

\setlength{\topmargin}{-0.5cm}
\setlength{\textheight}{22cm}
\setlength{\oddsidemargin}{0cm}
\setlength{\evensidemargin}{0cm}
\setlength{\textwidth}{15.8cm}
\setlength{\headsep}{0in}
\setlength{\parskip}{.15in}
\newcommand{\sectionsize}{\fontsize{12}{16}\selectfont}
\renewcommand{\bibfont}{\large}

\onecolumngrid

\begin{center}
{\huge\sf {Supplemental Material}} \\
\vspace{0.4cm}
{\huge\sf {The Competition for Shortest Paths on Sparse Graphs}} \\
{\LARGE\sf {Chi Ho Yeung and David Saad}} \\
\end{center}

{\large

\section{\sectionsize Proof of the convexity of $E^V(\flow)$}

Given $\alpha\ge 1$, we will show that $E(\flow)$ is convex by a self-consistent argument. To show that $E^V_i(\flow_{il})$ is convex, it is sufficient to show that $\Delta^+(\flow_{il})\le \Delta^+(\flow_{il}+1)$ for all $\flow_{il}$, given the $E^V_j(\flow_{ji})$ functions of descendants, are convex. From Eq.~(6), we have
\begin{align}
\Delta^+_i(\flow_{il}+1) &= |\flow_{il}+2|^\alpha-|\flow_{il}+1|^\alpha+\min_j\{\Delta^+_j[\flow_{ji}^*(\flow_{il}+1)]\}
\nonumber\\
&\ge |\flow_{il}+2|^\alpha-|\flow_{il}+1|^\alpha+\min_j\{\Delta^+_j[\flow_{ji}^*(\flow_{il})], \Delta^+_j[\flow_{ji}^*(\flow_{il})+1]\}
\nonumber\\
&\qquad\qquad\qquad\qquad
(\because \flow_{ji}^*(\flow_{il}+1)=\flow_{ji}^*(\flow_{il}) \mbox{ or } \flow_{ji}^*(\flow_{il})+1, \mbox{from Eq. (6)})
\nonumber\\
&=|\flow_{il}+2|^\alpha-|\flow_{il}+1|^\alpha+\min_j\{\Delta^+_j[\flow_{ji}^*(\flow_{il})]\}
\nonumber\\
&
\qquad\qquad\qquad\qquad\qquad\qquad\qquad\qquad\qquad\qquad
(\because \mbox{convexity of $E^V_j(\flow_{ji})$})
\nonumber\\
&\ge |\flow_{il}+1|^\alpha-|\flow_{il}|^\alpha+\min_j\{\Delta^+_j[\flow_{ji}^*(\flow_{il})]\}
\nonumber\\
&
\qquad\qquad\qquad\qquad\qquad\qquad\qquad\qquad
\mbox{(equality holds only when $\alpha=1$)}
\nonumber\\
&= \Delta^+_i(\flow_{il}).
\nonumber
\end{align}
Therefore, given the convexity of descendent $E^V_j(\flow_{ji})$'s, we can show that $E^V_i(\flow_{il})$ is convex. Hence, $E^V(\flow)$ is convex by a self-consistent argument.

\section{\sectionsize Simplifying the Computation of $E^V(\flow)$}

To compute $E^V(\flow)$, we first note that one has to evaluate $E^V(\flow)$ on a finite range of $\flow$ since $E^V(\flow)$ cannot be expressed by simple parametrization. Fortunately, $-Nf_s\le\flow\le Nf_s$ as $Nf_s$ is the maximum possible occupancy on any link. We can then simplify the iteration of Eq.~(2) by using Eq.~(6) with the following procedures: (i) evaluate $\tilde\flow_{il}=\Lambda_i+\sum_{j\in\cN\backslash l}\hat\flow_{ji}$, where $\hat\flow_{ji} = \argmin_{\flow_{ji}}E^V_{j}(\flow_{ji})$, (ii) evaluate $E^V_{j}(\tilde\flow_{il}) = |\tilde\flow_{il}|^\alpha+\sum_{j\in\cN\backslash l}E^V_{j}(\hat\flow_{ji})$, (iii) calculate $E^V_i(\flow_{il})$ in the relevant range of $\flow_{il}$ by $\Delta_i^\pm$ increase or decrease of $\flow_{il}$ from $\tilde\flow_{il}$.
Another step which makes use of backward messages~\cite{wong06a} can greatly simplify the iteration of Eq.~(2) by limiting the relevant $\flow$ to a small range; this result will be presented elsewhere~\cite{yeung11}.

For solving $P[E^V]$ numerically from Eq.~(2), instead of having only one receiver as in the simulations of finite system, one has to assume a small fraction $f_r$ of receiver, to gauge between the infinite trees of the cavity method and the finite number of nodes/receivers studied in simulations. Such an assumption would make the cavity result slightly deviate from the finite system considered in simulations with only one receiver, since in addition to the choice of path, senders may as well choose to use different receivers. We found that this is the reason for the small offset between theory and simulations in Fig.~3(a) as the offset vanishes when we simulate larger systems with more than a single receiver.

\section{\sectionsize Identification of the RSB phase in cases with multiple receiver type}

Similar to the case with only a single type of receiver, Eq.~(7) allows us to characterize the RS-RSB transition in the case with multiple receiver types. Unlike the previous case, where $\delta\Delta_i$ depends on a single perturbation $\delta\Delta_j$, the present perturbation $\delta\Delta_i$ may depend on a number of descendants $\delta\Delta_j$'s as given by Eq.~(7). We denote this number of descendants as $n_i^{x, y}(\flow_{il}^A, \flow_{il}^B)$, which depends on the flow $\flow_{il}^A, \flow_{il}^B$ and the corresponding changes $x$ and $y$, for instance, $(+1,-1)$. To obtain $\avg{n}$, one may average $n_i$ uniformly over a range of $\flow_{il}^A, \flow_{il}^B$ and with any $x$ and $y$, but this may not be physical. Instead, we average $n_i$ over a distribution $P(\flow_{il}^{A*}, \flow_{il}^{B*}, x, y)$, where $\flow_{il}^{A*}$ and $\flow_{il}^{B*}$ are the ground state flows when $E^V_{i}$ is merged according to the topology and $(\flow_{il}^{A*}+x, \flow_{il}^{B*}+y)$ represent another state with zero or minimal energy increment from the ground state. In this case, $\avg{n}$ characterizes the stability of the ground state with respect to a degenerate or excited state, and $\avg{n}<1$ and $\avg{n}>1$ characterize the RS and RSB phases, respectively.

}


\begin{thebibliography}{0}
\bibitem{footnote} Alternatively, in the large system limit, one may select a small fraction of the system nodes $f_r$ to act as receivers, rather than selecting a single receiver.
\end{thebibliography}
\end{document}